\documentclass[conference]{IEEEtran}
\IEEEoverridecommandlockouts
\usepackage{cite}
\usepackage{amsmath,amssymb,amsfonts,amsthm}
\usepackage[all=normal,bibbreaks=tight,paragraphs=tight,floats=tight,
mathspacing=tight,wordspacing=tight,charwidths=tight,leading=tight,
indent=tight,lists=tight]{savetrees}
\usepackage{algorithmic}
\usepackage{enumitem}
\usepackage{graphicx}
\usepackage{caption}
\usepackage{subcaption}
\usepackage{textcomp}
\usepackage{flushend}
\def\BibTeX{{\rm B\kern-.05em{\sc i\kern-.025em b}\kern-.08em
    T\kern-.1667em\lower.7ex\hbox{E}\kern-.125emX}}

\newtheorem{theorem}{Theorem}
\newtheorem{lemma}{Lemma}
    
\begin{document}

\setlength{\textfloatsep}{1\baselineskip}
\setlength{\belowcaptionskip}{-5pt}
%\setlength{\abovecaptionskip}{-3pt}

%\makeatletter
%\let\origsection\section
%\renewcommand\section{\@ifstar{\starsection}{\nostarsection}}
%
%\newcommand\nostarsection[1]
%{\sectionprelude\origsection{#1}\sectionpostlude}
%\newcommand\starsection[1]
%{\sectionprelude\origsection*{#1}\sectionpostlude}
%
%\newcommand\sectionprelude{%
%  \vspace{-0.3em}
%}
%
%\newcommand\sectionpostlude{%
%  \vspace{-0.2em}
%}
%\makeatother
%
%\makeatletter
%\let\origsubsection\subsection
%\renewcommand\subsection{\@ifstar{\starsubsection}{\nostarsubsection}}
%
%\newcommand\nostarsubsection[1]
%{\subsectionprelude\origsubsection{#1}\subsectionpostlude}
%
%\newcommand\starsubsection[1]
%{\subsectionprelude\origsubsection*{#1}\subsectionpostlude}
%
%\newcommand\subsectionprelude{%
%  \vspace{-0.3em}
%}
%
%\newcommand\subsectionpostlude{%
%  \vspace{-0.2em}
%}
%\makeatother

\title{Bias-variance tradeoff in MIMO channel estimation %\\ channel estimation with a physical model
\thanks{\tiny
This work has been performed in the framework of the Horizon 2020 project ONE5G (ICT-760809) receiving funds from the European Union. The authors would like to acknowledge the contributions of their colleagues in the project, although the views expressed in this contribution are those of the authors and do not necessarily represent the project.}
}

\author{\IEEEauthorblockN{Luc Le Magoarou, St\'ephane Paquelet}
\IEEEauthorblockA{\textit{b\raisebox{0.2mm}{\scalebox{0.7}{\textbf{$<>$}}}com}, Rennes, France}
}

\maketitle

\begin{abstract}
Channel estimation is challenging in multi-antenna communication systems, because of the large number of parameters to estimate. It is possible to facilitate this task by using a physical model describing the multiple paths constituting the channel, in the hope of reducing the number of unknowns in the problem. Adjusting the number of estimated paths leads to a bias-variance tradeoff. This paper explores this tradeoff, aiming to find the optimal number of paths to estimate. Moreover, the approach based on a physical model is compared to the classical least squares and Bayesian techniques. Finally, the impact of channel estimation error on the system data rate is assessed. 
\end{abstract}


\section{Introduction}

Multiple-input multiple-output (MIMO) communication systems allow for a dramatic increase in channel capacity, adding space to the classical time and frequency dimensions \cite{Telatar1999,Tse2005}. This is done by using several antennas at the transmitter ($N_t$) and at the receiver ($N_r$). The capacity of MIMO systems is maximized if the channel state is perfectly known at both ends of the link.% gains over single antenna systems are at most proportional to $\min(N_r,N_t)$.%, provided signal processing is done correctly.

%Maximal capacity is attained if the channel state is perfectly known by both the transmitter and the receiver, which requires to estimate the channel. 
Channel estimation is deeply impacted by the transition from single antenna to MIMO systems. Indeed, it amounts to determine a complex gain for each transmit/receive antenna pair, the narrowband (single carrier) MIMO channel as a whole being usually represented as a matrix $\mathbf{H} \in \mathbb{C}^{N_r \times N_t}$ of such gains. The number of real parameters to estimate is thus $2N_rN_t$, which may be very large for \emph{massive MIMO} systems, i.e.\ systems with up to several hundreds of antennas \cite{Rusek2013,Larsson2014}.

In a massive MIMO context, the classical least squares (LS) estimator is not adapted because the high dimensionality of the parameter space leads to an ill-posed problem. Therefore, in order to add prior information, it has been proposed classically to use Bayesian estimation and thus model the channel matrix as random, giving rise to estimators such as the linear minimum mean square error (LMMSE) \cite{Bjornson2017, Bazzi2017}. Another possibility is to use a parametric model based on the physics of wave propagation, in which the channel is expressed as a sum of $p$ paths \cite{Sayeed2002}. Whereas LS and LMMSE estimators have been studied extensively in terms of optimal training sequences and performance \cite{Biguesh2006}, a similar study is still lacking for channel estimators based on a physical model. 

\noindent {\bf Contributions.} In this paper, the performance of MIMO channel estimators based on a physical model is studied, highlighting a bias-variance tradeoff driven by the number of paths $p$ considered by the model. Moreover, physical channel estimators are compared to classical ones in a massive MIMO context. %(to do so, the results of \cite{Biguesh2006} are briefly adapted to fit the massive MIMO framework). 
Finally, the impact of channel estimation error on the system data rate is theoretically assessed.% The remainder of the paper is organized as follows. First, after having formulated the problem, the results of \cite{Biguesh2006} are briefly restated to fit the massive MIMO framework. Then, the physical channel model is studied in terms of bias and variance...\cite{Lemagoarou2018}

\section{Problem formulation}
\label{sec:problem_formulation}

\noindent{\bf Notations.} Matrices and vectors are denoted by bold upper-case and lower-case letters: $\mathbf{A}$ and $\mathbf{a}$ (except 3D ``\emph{spatial}'' vectors that are denoted $\overrightarrow{a}$); the $i$th column of a matrix $\mathbf{A}$ by $\mathbf{a}_i$; its entry at the $i$th line and $j$th column by $a_{ij}$. %The $i$th row of a matrix $\mathbf{A}$ is denoted by: $\mathbf{a}^i$ (it is a column vector).
A matrix transpose, conjugate and transconjugate is denoted by $\mathbf{A}^T$, $\mathbf{A}^*$ and $\mathbf{A}^H$ respectively. The trace of a linear transformation represented by $\mathbf{A}$ is denoted $\text{Tr}(\mathbf{A})$.
%Sets are denoted by calligraphic symbols: $\mathcal{A}$.
For matrices $\mathbf{A}$ and $\mathbf{B}$, $\mathbf{A}\geq \mathbf{B}$ means that $\mathbf{A}-\mathbf{B}$ is positive semidefinite.
 The linear span of a set of vectors $\mathcal{A}$ is denoted: $\text{span}(\mathcal{A})$. The Kronecker product and vectorization operators are denoted by $\otimes$ and $\text{vec}(\cdot)$ respectively. The identity matrix is denoted by $\mathbf{Id}$. $\mathcal{CN}(\boldsymbol\mu,\boldsymbol{\Sigma})$ denotes the standard complex gaussian distribution with mean $\boldsymbol\mu$ and covariance $\boldsymbol{\Sigma}$. $\mathbb{E}(\cdot)$ denotes expectation and $\text{cov}(\cdot)$ the covariance of its argument. 

\subsection{Channel estimation}
\label{ssec:channel_estimation}
Consider a narrowband block fading channel between a transmitter and a receiver with respectively $N_t$ and $N_r$ antennas. It is represented by the matrix $\mathbf{H} \in \mathbb{C}^{N_r \times N_t}$, in which $h_{ij}$ corresponds to the channel between the $j$th transmit and $i$th receive antennas.

In order to carry out channel estimation, $N_s$ known pilot symbols are sent through the channel by each transmit antenna. The corresponding training matrix is denoted $\mathbf{X} \in \mathbb{C}^{N_t \times N_s}$. It obeys a power constraint of the form $\left\Vert \mathbf{X} \right\Vert_F^2 = \mathcal{P}$ where $\mathcal{P}$ is the total training energy. In addition, the transmit power is assumed constant during the training phase, leading to $\left\Vert \mathbf{x}_i \right\Vert_2^2 = \mathcal{P}_e,\, i=1,\dots,N_s$, which implies $\mathcal{P} = N_s\mathcal{P}_e$. The signal at the receive antennas is thus expressed as $\mathbf{HX} + \mathbf{N}$, where $\mathbf{N}$ is a noise matrix (usually $\text{vec}(\mathbf{N})\sim \mathcal{CN}(0,\sigma^2\mathbf{Id})$). One useful quantity is the
potential signal to noise ratio (pSNR) defined as 
\begin{equation}
\text{pSNR}\triangleq  \frac{\mathcal{P}_e\left\Vert \mathbf{H} \right\Vert_F^2}{\sigma^2}.
\label{eq:pSNR}
\end{equation}
It is an upper bound on the signal to noise ratio (SNR), which is used to quantify the performance of channel estimators in section~\ref{sec:perf_of_estimators}. Note that pSNR and SNR coincide for a rank one channel combined with an optimal precoding at the transmitter, and that with no precoding at all, $\text{SNR} = \text{pSNR} - \log_{10}(N_t)$ (in dB).
 Due to the high cost and power consumption of millimeter wave Radio Frequency (RF) chains, it has been proposed to have less RF chains than antennas in both the transmitter and receiver \cite{Elayach2014,Heath2016,Sayeed2016}. Such systems are often referred to as \emph{hybrid architectures}. Mathematically speaking, this translates into sensing the channel through analog precoders $\mathbf{v}_i \in \mathbb{C}^{N_t}$, $i=1,\dots,N_{\text{RF}}$, with $N_{\text{RF}}\leq N_t$, as well as observing the signal at the receiver through analog combiners denoted $\mathbf{w}_j \in \mathbb{C}^{N_r}$, $j=1,\dots,N_c$. The observed data is thus expressed in all generality as
\begin{equation}
\mathbf{Y} = \mathbf{W}^H\mathbf{H}\mathbf{X} + \mathbf{W}^H\mathbf{N},
\label{eq:observation_model}
\end{equation}
where $\mathbf{W} \triangleq (\mathbf{w}_1,\dots,\mathbf{w}_{N_c})$ and the training matrix is constrained to be of the form $\mathbf{X} = \mathbf{VZ}$, where $\mathbf{Z} \in \mathbb{C}^{N_{\text{RF}} \times N_{s}}$ is the digital training matrix, and $\mathbf{V} \in \mathbb{C}^{N_{\text{t}} \times N_{\text{RF}}}$ is the analog precoding matrix. %Channel estimation amounts to retrieve $\mathbf{H}$, knowing $\mathbf{Y}$, $\mathbf{X}$, $\mathbf{W}$ and the noise distribution. In case of a parametric model, as we consider in this paper, retrieving $\mathbf{H}$ is equivalent to retrieving the parameters $\boldsymbol{\theta}$.
For more convenience and to lighten notations, one can introduce the equivalent vectorized observation model:
\begin{equation}
\mathbf{y} = \mathbf{M}\mathbf{h} + \mathbf{n},
\label{eq:observation_model_vect}
\end{equation}
where $\mathbf{y} \triangleq \text{vec}(\mathbf{Y})\in \mathbb{C}^{N_cN_s}$, $\mathbf{h} \triangleq \text{vec}(\mathbf{H})\in \mathbb{C}^{N_rN_t}$, $\mathbf{n} \triangleq \text{vec}(\mathbf{W}^H\mathbf{N})\in \mathbb{C}^{N_cN_s}$ and $\mathbf{M} \triangleq \mathbf{X}^T\otimes\mathbf{W}^H\in \mathbb{C}^{N_cN_s \times N_rN_t}$ is called the \emph{observation matrix}. In the remaining of the paper, it is assumed that the analog combiners are mutually orthogonal and of unit norm: $\mathbf{w}_i^H\mathbf{w}_j = \delta_{ij}$. This yields $\left\Vert \mathbf{W} \right\Vert_F^2 = N_c$ and $\mathbf{n} \sim \mathcal{CN}(0,\sigma^2\mathbf{Id})$.

\noindent {\bf Objective.} Channel estimation aims at retrieving $\mathbf{h}$ (or equivalently $\mathbf{H}$) from the observation of $\mathbf{y}$, knowing the observation matrix $\mathbf{M}$ and the distribution of the noise vector $\mathbf{n}$. The channel estimator is denoted $\hat{\mathbf{h}}$ (or equivalently $\hat{\mathbf{H}}$ in matrix form).
At first sight, the search space of channel estimation is thus of dimension $2N_rN_t$, which may be very large in massive MIMO systems (up to several thousands). For this reason, classical estimation methods such as the least squares (LS) may not be appropriate. In order to overcome this limitation, some information about the channel has to be used to regularize the problem. For example, the channel can be considered as a random vector whose distribution is known, yielding Bayesian channel estimation \cite{Bjornson2017, Bazzi2017}. Another possibility is to use a physical channel model, as presented in the next subsection.

\subsection{Physical channel model}
\label{ssec:physical_channel_model}

Inspired by the physics of propagation under the plane waves assumption, it has been proposed to express the channel as a sum of rank one matrices, each corresponding to a single physical path between transmitter and receiver. This approach is used by most MIMO channel simulators \cite{Sun2017,Jaeckel2014,3GPP2017} and has been validated by propagation measurements. Adopting this kind of modeling and generalizing it to take into account any three-dimensional antenna array geometry, channel matrices take the form
\begin{equation}
\mathbf{H} =  \sum\nolimits_{i=1}^Pc_i\mathbf{e}_r(\overrightarrow{u_{r,i}})\mathbf{e}_t(\overrightarrow{u_{t,i}})^H,
\label{eq:channel_model_phys}
\end{equation}
where $P$ is the total number of considered \emph{physical paths} (up to a few hundreds in classical simulators), $c_i \triangleq \rho_i \mathrm{e}^{\mathrm{j}\phi_i}$ is the complex gain of the $i$th path, $\overrightarrow{u_{t,i}}$ is the unit vector corresponding to its direction of departure (DoD) and $\overrightarrow{u_{r,i}}$ the unit vector corresponding to its direction of arrival (DoA)(both being described in spherical coordinates by an azimuth angle $\eta$ and an elevation angle $\psi$). The steering vectors $\mathbf{e}_r(\overrightarrow{u})\in \mathbb{C}^{N_r}$ and $\mathbf{e}_t(\overrightarrow{u})\in \mathbb{C}^{N_t}$ are defined as $(\mathbf{e}_x(\overrightarrow{u}))_j = \frac{1}{\sqrt{N_x}} \mathrm{e}^{-\mathrm{j}\frac{2\pi}{\lambda}\overrightarrow{a_{x,j}}.\overrightarrow{u}}$ for $x\in \{r,t\}$. 
%$\mathbf{e}_r(\overrightarrow{u}) \triangleq \frac{1}{\sqrt{N_r}} (\mathrm{e}^{-\mathrm{j}\frac{2\pi}{\lambda}\overrightarrow{a_{r,1}}.\overrightarrow{u}},\dots,\mathrm{e}^{-\mathrm{j}\frac{2\pi}{\lambda}\overrightarrow{a_{r,N_r}}.\overrightarrow{u}})^T$ and $\mathbf{e}_t(\overrightarrow{u}) \triangleq \frac{1}{\sqrt{N_t}} (\mathrm{e}^{-\mathrm{j}\frac{2\pi}{\lambda}\overrightarrow{a_{t,1}}.\overrightarrow{u}},\dots,\mathrm{e}^{-\mathrm{j}\frac{2\pi}{\lambda}\overrightarrow{a_{t,N_t}}.\overrightarrow{u}})^T$
The set $\mathcal{A}_x\triangleq\{\overrightarrow{a_{x,1}},\dots,\overrightarrow{a_{x,N_x}}\}$ gathers the positions of the antennas with respect to the centroid of the array (transmit if $x=t$, receive if $x=r$). %In order to lighten notations, the matrix $\mathbf{A}_x \triangleq \frac{2\pi}{\lambda}(\overrightarrow{a_{x,1}},\dots,\overrightarrow{a_{x,n_x}}) \in\mathbb{R}^{3\times N_x}$ is introduced. It simplifies the steering/response vector expression to $\mathbf{e}_x(\overrightarrow{u}) = \frac{1}{\sqrt{n_x}} \mathrm{e}^{-\mathrm{j}\mathbf{A}_x^T\overrightarrow{u}}$, where the exponential function is applied component-wise.
The same model is used in \cite{Lemagoarou2018}. 
%In order to further lighten notations, the $p$th \emph{atomic channel} is defined as $\mathbf{H}_p \triangleq c_p\mathbf{e}_r(\overrightarrow{u_{r,p}}).\mathbf{e}_t(\overrightarrow{u_{t,p}})^H$, and its vectorized version $\mathbf{h}_p \triangleq \text{vec}(\mathbf{H}_p) \in \mathbb{C}^{N_rN_t}$. Therefore, defining the vectorized channel $\mathbf{h} \triangleq \text{vec}(\mathbf{H})$, yields $\mathbf{h} = \sum_{p=1}^P \mathbf{h}_p$. 
%Note that the channel description used here is very general, as it handles any three-dimensional antenna array geometry, and not only Uniform Linear Arrays (ULA) or Uniform Planar Arrays (UPA) as is sometimes proposed.

%\begin{equation}
%\mathbf{H} = \sum_{q=1}^Q\sum_{p\in \mathcal{C}_q}c_p\mathbf{e}_r(\overrightarrow{u_{r,p}}).\mathbf{e}_t(\overrightarrow{u_{t,p}})^H
%\label{eq:channel_model_cluster}
%\end{equation}

%In such physical channel models, it is classically assumed that paths are grouped into $Q$ clusters, in which the DoDs and DoAs are close \cite{Saleh1987, Wallace2002, Jensen2004}. This corresponds to a partition of the index set into $Q$ subsets: $\{1,\dots,P\} = \bigcup_{q=1}^Q \mathcal{C}_q$ with $\mathcal{C}_{q_1} \cap \mathcal{C}_{q_2} = \emptyset$ if $q_1 \neq q_2$.

\noindent {\bf Estimation model.} It makes sense to have the channel estimator $\hat{\mathbf{H}}$ take a form similar to that of the physical channel $\mathbf{H}$ given in \eqref{eq:channel_model_phys}. It has indeed been proposed in several contexts \cite{Bajwa2010,Alkhateeb2014,Venugopal2017}, leading to estimators expressed as
\begin{equation}
\hat{\mathbf{H}} =  \sum\nolimits_{k=1}^pd_k\mathbf{e}_r(\overrightarrow{v_{r,k}})\mathbf{e}_t(\overrightarrow{v_{t,k}})^H,
\label{eq:channel_model_virt}
\end{equation}
where $p$ is the number of paths considered for estimation, called \emph{virtual paths}. In practice, one takes $p\ll P$ (with $p$ at most a few dozens), because several physical paths can be merged into a single virtual path without harming a lot description accuracy. Indeed, steering vectors associated to close enough directions are almost collinear, yielding a limited angular resolution for the system \cite{Sayeed2002}. Moreover, a smaller $p$ leads to a better conditioned problem, as will be evidenced in the next section. Let us define the \emph{model set} $\mathcal{M}_p \triangleq \big\{ \mathbf{A} \in \mathbb{C}^{N_r\times N_t},\, \mathbf{A} = \sum\nolimits_{k=1}^p  d_k\mathbf{e}_r(\overrightarrow{v_{r,k}})\mathbf{e}_t(\overrightarrow{v_{t,k}})^H \big\} $ to which the estimator made of $p$ virtual paths belongs. By abuse of notation, the vectorized version of this set will be denoted the same way, such that $\mathcal{M}_p=\big\{ \mathbf{a} \in \mathbb{C}^{N_r N_t},\, \mathbf{a} = \sum\nolimits_{k=1}^p  d_k\mathbf{e}_t(\overrightarrow{v_{t,k}})^*\otimes\mathbf{e}_r(\overrightarrow{v_{r,k}}) \big\} $. Such sets obey the inclusion relation $\mathcal{M}_{q} \subset \mathcal{M}_{q+1}$.

The estimation model can be seen as a parametric model with parameter vector $\boldsymbol{\theta} = \{\boldsymbol{\theta}^{(k)}\triangleq(\rho_k,\phi_k,\eta_{r,k},\psi_{r,k},\eta_{t,k},\psi_{t,k}),\; k=1{\scriptstyle,\dots,}p\}$. There are thus $6p$ real parameters in this model (the complex gain, DoD and DoA of every path are described with two parameters each). Of course, the model is most useful for estimation in the case where $6p\ll 2N_rN_t$, since the number of parameters is thus greatly reduced. Conversely, taking $p$ too small may harm the descriptive power of the model. The number of virtual paths $p$ thus drives a tradeoff between complexity and expressiveness, whose study is the object of the next sections.

\section{Performance of estimators}
\label{sec:perf_of_estimators}

%\subsection{Relative mean square error}
Let us consider an estimator $\hat{\mathbf{h}}$ of a channel $\mathbf{h}$. Its performance is assessed using the relative mean square error (rMSE) defined as
\begin{equation}
\text{rMSE}(\hat{\mathbf{h}}) \triangleq \mathbb{E}\left[ \frac{\big\Vert \mathbf{h} - \hat{\mathbf{h}} \big\Vert_2^2}{\left\Vert \mathbf{h}  \right\Vert_2^2} \right].
\label{eq:rMSE}
\end{equation}
For any estimator, this error can be decomposed \cite{Kay1993} as
\begin{equation}
\text{rMSE}(\hat{\mathbf{h}}) = 
\frac{\big\Vert\mathbf{h}-\mathbb{E}[\hat{\mathbf{h}}]\big\Vert_2^2}{\left\Vert \mathbf{h} \right\Vert_2^2}
+
\frac{\text{Tr}[\text{cov}(\hat{\mathbf{h}})]}{\left\Vert \mathbf{h} \right\Vert_2^2},
\label{eq:bias_variance}
\end{equation}
where the first term called \emph{bias} represents the error due to the deviation of the average estimate from the true channel and the second called \emph{variance} represents the error due to fluctuations around the average. In this section and the following, the rMSE of various channel estimators is studied. The emphasis is put on physical channel estimators, for which a bias-variance tradeoff driven by the number of virtual paths $p$ is exhibited. The study of this tradeoff is made possible by the definition of an appropriate oracle estimator.

\subsection{Classical channel estimators}
\label{ssec:classical_estimators}
Let us now review the performance of the most classical MIMO channel estimators that are the least squares (LS) and the linear minimum mean square error (LMMSE). This subsection essentially summarizes the main results of \cite{Biguesh2006}, in a way that takes into account hybrid systems. %The two considered estimators are both \emph{unbiased}, meaning that the bias term of their rMSE is null. 

\noindent {\bf LS.}
The least squares estimator is an \emph{unbiased} estimator defined (using the vectorized notation of \eqref{eq:observation_model_vect}) as
\begin{equation}
\hat{\mathbf{h}}_{\text{LS}} \triangleq \underset{\mathbf{g}}{\text{argmin}} \left\Vert \mathbf{y} - \mathbf{Mg} \right\Vert_2^2 = (\mathbf{M}^H\mathbf{M})^{-1}\mathbf{M}^H\mathbf{y}.
\end{equation}
It requires to send $N_s\geq N_t$ pilot symbols to exist (for the matrix $\mathbf{M}^H\mathbf{M}$ to be invertible). In that case, the variance (and thus the rMSE) is minimized if
the observation matrix obeys the condition
\begin{equation}
\left(\mathbf{M}^H\mathbf{M}\right)_{\text{opt}} = \mathcal{P}_e \frac{N_cN_s}{N_rN_t}\mathbf{Id},
\end{equation}
leading to the optimal performance
\begin{equation}
\text{rMSE}_{\text{opt}}(\hat{\mathbf{h}}_{\text{LS}}) =    \frac{N_rN_t}{\text{pSNR}}\frac{N_rN_t}{N_cN_s}.
\label{eq:perf_LS}
\end{equation}
Note that in the case where $N_s=N_t$ (which corresponds to the minimum number of pilot symbols) and $N_c=N_r$ (which corresponds to a non-hybrid receiver), the optimal rMSE is proportional to $N_rN_t$, showing that such an estimator is not adapted to massive MIMO systems in which this quantity is large.

\noindent {\bf LMMSE.} The linear minimum mean square error estimator is a Bayesian estimator, meaning that the channel is assumed to be random and to follow a known distribution $\mathbf{h} \sim \mathcal{CN}(\mathbf{0},\mathbf{R})$. A channel realization can then be seen as drawing at random a user location in a given region in a fixed environment: the smaller the region the closer to singular $\mathbf{R}$. Note that in practice the covariance matrix $\mathbf{R}$ has to be estimated using previous channel estimates.%, making the implicit assumption that the channel has not changed a lot in the time interval used to estimate it. 
 If $\mathbf{R}$ is perfectly known, the LMMSE is the linear estimator minimizing the MSE, it takes the form
\begin{equation}
\hat{\mathbf{h}}_{\text{LMMSE}} \triangleq \mathbf{RM}^H\left( \mathbf{MRM}^H + \sigma^2\mathbf{Id} \right)^{-1}\mathbf{y},
\end{equation}
with
\begin{equation*}
\mathbf{RM}^H\left( \mathbf{MRM}^H + \sigma^2\mathbf{Id} \right)^{-1} = \underset{\mathbf{A}}{\text{argmin}}~\,\, \mathbb{E}\left[ \left\Vert\mathbf{h} - \mathbf{Ay}\right\Vert_2^2\right] ,
\end{equation*}
where the expectation is taken over both the noise and the channel distributions.
In that case, the rMSE is minimized if
the observation matrix obeys the condition (for a high enough SNR, see \cite[section V]{Biguesh2006} for more details)
\begin{equation}
\left(\mathbf{M}^H\mathbf{M}\right)_{\text{opt}} = \left(\mathcal{P}_e \frac{N_cN_s}{N_rN_t} + \frac{\sigma^2}{N_rN_t}\text{Tr}\left[\mathbf{R}^{-1}\right]\right)\mathbf{Id} - \sigma^2\mathbf{R}^{-1}.
\end{equation}
This leads to the optimal performance
\begin{equation}
\text{rMSE}_{\text{opt}}(\hat{\mathbf{h}}_{\text{LMMSE}}) =    \frac{1}{\frac{\mathbb{E}[\text{pSNR}]}{N_rN_t}\frac{N_cN_s}{N_rN_t} + \frac{\text{Tr}\left[\mathbf{R}^{-1}\right]\text{Tr}\left[\mathbf{R}\right]}{(N_rN_t)^2}}.
\label{eq:perf_LMMSE}
\end{equation}
In this expression, the first term at the denominator $\frac{\mathbb{E}[\text{pSNR}]}{N_rN_t}\frac{N_cN_s}{N_rN_t}$ corresponds to the expected inverse of the optimal LS performance, indicating that the LMMSE should exhibit the same behaviour for massive MIMO systems. However, the second term $\frac{\text{Tr}\left[\mathbf{R}^{-1}\right]\text{Tr}\left[\mathbf{R}\right]}{(N_rN_t)^2}$ may compensate for the first one. This term is lower-bounded by one (if all the eigenvalues of $\mathbf{R}$ are equal), and grows as the eigenvalues distribution becomes more uneven (if $\mathbf{R}$ is closer to be singular). Simply put, the distribution of the eigenvalues of $\mathbf{R}$ determines to which extent knowing it is informative for channel estimation. The covariance being estimated in practice with previous channel estimates, the smaller the used time interval, the closer to singular $\mathbf{R}$ (since the user stayed in a smaller region), so that it gives more information and the optimal rMSE is reduced. %Note that in practical systems, the covariance matrix $\mathbf{R}$ needs to be estimated, which poses another estimation problem, source of potential errors.

\subsection{Physical channel estimators}
As seen in the previous subsection, the LS and LMMSE estimators have been theoretically studied, and their performance is well understood. In contrast, estimators based on a physical model taking the form of \eqref{eq:channel_model_virt} have not been investigated as deeply. This may be due to the difficulty to study jointly the bias and variance terms of the rMSE. To overcome this difficulty, an oracle estimator is introduced here which allows to separate the analysis of the bias from that of the variance. This oracle should be seen as a tool to study theoretically estimators based on a physical model, and its relevance with respect to practical estimators is assessed in section~\ref{sec:experiments}. 

\noindent {\bf Oracle definition.} First, one can notice that for an estimator $\hat{\mathbf{h}} \in \mathcal{M}_p$,
$$
\text{rMSE}(\hat{\mathbf{h}}) \geq \frac{\big\Vert\mathbf{h}-\text{proj}_{\mathcal{M}_p}(\mathbf{h})\big\Vert_2^2}{\left\Vert \mathbf{h} \right\Vert_2^2}.
$$
where $\text{proj}_{\mathcal{M}_p}(\mathbf{u}) \triangleq \text{argmin}_{\mathbf{a}\in \mathcal{M}_p}\, \left\Vert \mathbf{u} - \mathbf{a} \right\Vert.$
Then, the oracle estimator $\hat{\mathbf{h}}_{\text{oracle}}$ is defined as the \emph{unbiased efficient} estimator of $\text{proj}_{\mathcal{M}_p}(\mathbf{h})$. % considering an observation model modified with respect to \eqref{eq:observation_model_vect}, as 
%\begin{equation}
%\tilde{\mathbf{y}} = \mathbf{M}\text{proj}_{\mathcal{M}_p}(\mathbf{h}) +\mathbf{n}.
%\end{equation}
 In other words, the oracle property amounts to consider that $\mathbf{M}\text{proj}_{\mathcal{M}_p}(\mathbf{h}) + \mathbf{n}$ is observed instead of $\mathbf{M}\mathbf{h} + \mathbf{n}$, and thus ignore the part of the channel that is orthogonal to the model set $\mathcal{M}_p$. %suppress from the observation the part of the channel that cannot be expressed by the model $\mathcal{M}_p$.
  Mathematically speaking, using the oracle allows to study separately the bias and variance terms of the rMSE, as done below. Intuitively, the variance should increase with $p$ whereas the bias should decrease, since $p$ represents the complexity/expressiveness of the model.

\noindent {\bf Variance.} The oracle being \emph{efficient} means that
$$
\text{Tr}[\text{cov}(\hat{\mathbf{h}}_{\text{oracle}})] =  \text{CRB}
$$ where CRB is the Cram\'er-Rao lower bound \cite{Rao1945,Cramer1946}. This bound has been derived in \cite{Lemagoarou2018}, leading to the following bound for the variance term of \eqref{eq:bias_variance}.

\begin{theorem}
The variance of the oracle estimator is given by
\begin{equation}
\frac{\text{\emph{Tr}}[\text{\emph{cov}}(\hat{\mathbf{h}}_{\text{\emph{oracle}}})]}{\left\Vert \mathbf{h} \right\Vert_2^2}\geq \frac{3p}{\text{\emph{pSNR}}}.
\label{eq:thm_variance}
\end{equation}
\end{theorem}
The important feature of this result is the fact that the bound on the variance is proportional to the number of virtual paths $p$ (see \cite{Lemagoarou2018} for the proof of the theorem). Moreover, denoting
\begin{equation}
\text{proj}_{\mathcal{M}_p}(\mathbf{h}) = \sum\nolimits_{i=1}^pb_i\mathbf{e}_t(\overrightarrow{w_{t,i}})^*\otimes\mathbf{e}_r(\overrightarrow{w_{r,i}}),
\label{eq:channel_model_oracle}
\end{equation}
the inequality is replaced by an equality in the theorem if the observation matrix $\mathbf{M} =\mathbf{X}^T\otimes\mathbf{W}^H$ obeys the two conditions
$$
\text{span}\left(\bigcup\nolimits_{i=1}^p\Big\{\mathbf{e}_t(\overrightarrow{w_{t,i}}),\tfrac{\partial\mathbf{e}_t(\overrightarrow{w_{t,i}})}{\partial\eta_{t,i}},\tfrac{\partial\mathbf{e}_t(\overrightarrow{w_{t,i}})}{\partial\psi_{t,i}}\Big\}\right) \subset \text{im}(\mathbf{X}),
$$
$$
\text{span}\left(\bigcup\nolimits_{i=1}^p\Big\{\mathbf{e}_r(\overrightarrow{w_{r,i}}),\tfrac{\partial\mathbf{e}_r(\overrightarrow{w_{r,i}})}{\partial\eta_{r,i}},\tfrac{\partial\mathbf{e}_r(\overrightarrow{w_{r,i}})}{\partial\psi_{r,i}}\Big\}\right) \subset \text{im}(\mathbf{W}).
$$
Note that the variance is in that case directly proportional to the number of parameters to estimate. Moreover the result is pretty general. Indeed, for example if the receiver (or transmitter) has only one antenna, the theorem remains valid with the right-hand side of \eqref{eq:thm_variance} becoming $\frac{2p}{\text{pSNR}}
$, since the two parameters corresponding to the DoA (or DoD) of each virtual path disappear, leading to only four parameters per path.

\noindent {\bf Bias.} The oracle being \emph{unbiased} with respect to $\text{proj}_{\mathcal{M}_p}(\mathbf{h})$ means that 
$
\mathbb{E}[\hat{\mathbf{h}}_{\text{oracle}}]  = \text{proj}_{\mathcal{M}_p}(\mathbf{h}).
$
In that case, the bias term of \eqref{eq:bias_variance} becomes
\begin{equation}
\frac{\big\Vert\mathbf{h}-\mathbb{E}[\hat{\mathbf{h}}_{\text{oracle}}]\big\Vert_2^2}{\left\Vert \mathbf{h} \right\Vert_2^2} = \frac{\big\Vert\mathbf{h}-\text{proj}_{\mathcal{M}_p}(\mathbf{h})\big\Vert_2^2}{\left\Vert \mathbf{h} \right\Vert_2^2}.
\label{eq:bias_oracle}
\end{equation}
The study of this bias term is the main contribution of this paper. It is carried out in the next section.

\section{Bias of physical channel estimators}
\label{sec:bound_bias}

Computing the bias of the oracle estimator $\hat{\mathbf{h}}_{\text{oracle}}$ defined above amounts to compute the projection $\text{proj}_{\mathcal{M}_p}(\mathbf{h})$. Unfortunately, even considering a discretized set of candidates DoAs and DoDs, this problem (which then becomes a sparse approximation problem) is NP-hard \cite{Tropp2010}. The projection can be approximated numerically using sparse recovery methods, as will be done in section~\ref{sec:experiments}. However the objective of this section is to study theoretically the bias and give an interpretable upper bound. 

\noindent {\bf Assumptions.} Let us consider in this section the simpler case in which the transmitter or receiver has only one antenna ($N_t=1$ or $N_r=1$). In that case, the physical channel $\mathbf{h}$ and its estimator $\hat{\mathbf{h}} \in \mathcal{M}_p$ are expressed
$$
\mathbf{h} = \sum\nolimits_{i=1}^Pc_i\mathbf{e}(\overrightarrow{u_{i}}),\quad
\hat{\mathbf{h}} = \sum\nolimits_{k=1}^pd_k\mathbf{e}(\overrightarrow{v_{k}}),
$$
%and the estimator $\hat{\mathbf{h}} \in \mathcal{M}_p$ is expressed
where the indices denoting transmitter or receiver have been dropped since they are useless (they will be dropped in each notation defined in section~\ref{ssec:physical_channel_model} used in this section).
 This assumption allows to lighten considerably the notations in the following development. However, the general method used to bound the bias remains valid in the general model given by \eqref{eq:channel_model_phys} and \eqref{eq:channel_model_virt}.

\noindent {\bf Bound on the bias.} Starting from \eqref{eq:bias_oracle}, the inequality
$$
\big\Vert\mathbf{h}-\mathbb{E}[\hat{\mathbf{h}}_{\text{oracle}}]\big\Vert_2\leq \big\Vert\mathbf{h}-\hat{\mathbf{h}}\big\Vert_2
$$
is true for any $\hat{\mathbf{h}} \in \mathcal{M}_p$ (by definition of the projection). In particular, it is true when using the following suboptimal strategy to define $\hat{\mathbf{h}}$: 
\begin{enumerate}[leftmargin=*]
\item Partition the set of $P$ physical paths into $p$ subsets $\mathcal{R}_k, k=1,\dots,p$, of paths with close directions.
\item Assign one virtual path of direction $\overrightarrow{v_k}$ to each subset $\mathcal{R}_k$.
\item Set $d_k$ as the optimal coefficient to approximate the paths in each subset $\mathcal{R}_k$ by $d_k\mathbf{e}(\overrightarrow{v_k})$, given the direction $\overrightarrow{v_k}$.
\end{enumerate}
The two first steps of this strategy amount to write
$$
\mathbf{h}-\hat{\mathbf{h}} = \sum\nolimits_{k=1}^p\left[\sum\nolimits_{i\in \mathcal{R}_k} c_i\mathbf{e}(\overrightarrow{u_{i}}) - d_k\mathbf{e}(\overrightarrow{v_{k}})\right],
$$
and the last one yields $d_k = \sum\nolimits_{i\in \mathcal{R}_k} c_i\mathbf{e}(\overrightarrow{v_{k}})^H\mathbf{e}(\overrightarrow{u_{i}})$ (this is the coefficient of the orthogonal projection onto $\mathbf{e}(\overrightarrow{v_{k}})$). In that case, 
%$$
%\big\Vert\mathbf{h}-\hat{\mathbf{h}}\big\Vert_2= 
%\Big\Vert\sum\limits_{k=1}^p\Big[\sum\limits_{i\in \mathcal{R}_k} c_i\left(\mathbf{e}(\overrightarrow{u_{i}}) - \mathbf{e}(\overrightarrow{v_{k}})^H\mathbf{e}(\overrightarrow{u_{i}})\mathbf{e}(\overrightarrow{v_{k}})\right)\Big]\Big\Vert_2
%$$
%Applying the triangle inequality twice gives
\begin{align*}
\big\Vert\mathbf{h}-\hat{\mathbf{h}}\big\Vert_2&= 
\Big\Vert\sum\limits_{k=1}^p\Big[\sum\limits_{i\in \mathcal{R}_k} c_i\left(\mathbf{e}(\overrightarrow{u_{i}}) - \mathbf{e}(\overrightarrow{v_{k}})^H\mathbf{e}(\overrightarrow{u_{i}})\mathbf{e}(\overrightarrow{v_{k}})\right)\Big]\Big\Vert_2\\
&\leq 
\sum\limits_{k=1}^p\sum\limits_{i\in \mathcal{R}_k}\Big\Vert c_i\left(\mathbf{e}(\overrightarrow{u_{i}}) - \mathbf{e}(\overrightarrow{v_{k}})^H\mathbf{e}(\overrightarrow{u_{i}})\mathbf{e}(\overrightarrow{v_{k}})\right)\Big\Vert_2\\
&= 
\sum\limits_{k=1}^p\sum\limits_{i\in \mathcal{R}_k} |c_i|\sqrt{1 - |\mathbf{e}(\overrightarrow{v_{k}})^H\mathbf{e}(\overrightarrow{u_{i}})|^2},
\end{align*}
where the second line is obtained by applying the triangle inequality twice and the third by simply developing the norm in the second one.
At this point, the bias is bounded by a sum over all physical paths, each term being a quantity $\sqrt{1 - |\mathbf{e}(\overrightarrow{v_{k}})^H\mathbf{e}(\overrightarrow{u_{i}})|^2}$ measuring the non-collinearity of the steering vectors corresponding to the physical path and to the associated virtual path, weighted by the modulus of its physical coefficient $|c_i|$. To go further, let us analyze in details the scalar product between steering vectors appearing in the previous inequality. Its expression is given by
$$
\mathbf{e}(\overrightarrow{v_{k}})^H\mathbf{e}(\overrightarrow{u_{i}}) = \frac{1}{N}\sum\nolimits_{n=1}^N\mathrm{e}^{-\mathrm{j}\frac{2\pi}{\lambda}\overrightarrow{a_n}.(\overrightarrow{u_i} - \overrightarrow{v_k}) }.
$$
Using the infinite series representation of the exponential function leads to the following result.
\begin{lemma}
\label{lem:approx_scalar_product}
provided $\left\Vert\overrightarrow{u_i} - \overrightarrow{v_k} \right\Vert_2 < \frac{1}{\sqrt{2}\pi\left\Vert \frac{\overrightarrow{a_n}}{\lambda} \right\Vert_2 },\forall n$, 
\begin{align*}
&\sqrt{1 - |\mathbf{e}(\overrightarrow{v_{k}})^H\mathbf{e}(\overrightarrow{u_{i}})|^2}\leq\\ & 2\pi \left\Vert\overrightarrow{u_i} - \overrightarrow{v_k} \right\Vert_2 \sqrt{\frac{1}{N}\sum\nolimits_{n=1}^N\left\Vert \frac{\overrightarrow{a_n}}{\lambda} \right\Vert_2^2\cos^2\big(\overrightarrow{a_n},(\overrightarrow{u_i} - \overrightarrow{v_k})\big) }.
\end{align*}
\end{lemma}
The proof is given in appendix~\ref{app:proof_lemma}. If the condition of the lemma is fulfilled by each physical path (if there are enough virtual paths with appropriate directions), this yields 

\noindent\resizebox{\columnwidth}{!} 
{
\begin{minipage}{\columnwidth}
\begin{align*}
& \big\Vert\mathbf{h}-\mathbb{E}[\hat{\mathbf{h}}_{\text{oracle}}]\big\Vert_2 \leq \\
&\sum\limits_{k=1}^p\sum\limits_{i\in \mathcal{R}_k} 2\pi|c_i| \left\Vert\overrightarrow{u_i} - \overrightarrow{v_k} \right\Vert_2 \sqrt{\frac{1}{N}\sum_{n=1}^N\left\Vert \frac{\overrightarrow{a_n}}{\lambda} \right\Vert_2^2\cos^2\big(\overrightarrow{a_n},(\overrightarrow{u_i} - \overrightarrow{v_k})\big) }.
\end{align*}
\end{minipage}
}

%\begin{equation}
%\resizebox{.9 \textwidth}{!} 
%{
%    $ a + b $
%}
%\end{equation}

\noindent {\bf Interpretations.} %Several comments are in order regarding this bound. 
First, notice that the part under the square root represents the angular sensitivity of the antenna array: %it can be seen as the mean of the sensitivities of each antenna to a change in the direction of the difference $\overrightarrow{u_i} - \overrightarrow{v_k}$, which is proportional to the distance of the antenna to the array centroid (larger antenna arrays exhibit more angular sensitivity). 
it is bounded by the quantity $\kappa(\mathcal{A})\triangleq\sqrt{\frac{1}{N}\sum_{n=1}^N\big\Vert \frac{\overrightarrow{a_n}}{\lambda} \big\Vert_2^2}$%, which is attained if the difference of direction is aligned with the antenna array (for an ULA), and
which depends only on the antenna array geometry (larger antenna arrays exhibit more angular sensitivity). Second, let us assess the dependency of the bound to $p$. To do so, assume that the assignment of physical paths to virtual paths is done by partitioning a given portion of the sphere in $p$ regions and assigning all the physical paths whose direction falls into the $k$th region to the $k$th virtual path, whose direction is at the center of the region. That way, the bound is obviously decreasing when $p$ increases, since adding a virtual path reduces the distance $\left\Vert\overrightarrow{u_i} - \overrightarrow{v_k} \right\Vert_2$ for paths of the new region. The rate at which it decreases depends on the physical directions distribution. For example, in the worst-case scenario of uniformly distributed paths over a given portion of the sphere and regions of equal size, each of the $p$ regions gathers of the order of $\frac{1}{p}$ physical paths and the typical distance to the center of each region is of the order of $\frac{1}{\sqrt{p}}$. This leads to a bound of the order of $p\times \frac{1}{p} \times \frac{1}{\sqrt{p}}=\frac{1}{\sqrt{p}}$ and consequently to a bias in \eqref{eq:bias_oracle} of the order of $\frac{1}{p}$ at worst. The bias is evaluated numerically for realistic paths distributions in section~\ref{sec:experiments}.
 Finally, this bound lends itself to nice interpretations, but may be loose. A way of improving the bound is discussed in appendix~\ref{app:improving_the_bound}.

\section{Impact on the data rate} 
In order to assess the influence of channel estimation on the system performance, the rMSE in itself is of little use. Instead, the data rate loss caused by the estimation error (which leads to imperfect precoding) should be evaluated. To do so, assuming a transmission model of the form $\mathbf{z} = \mathbf{Hx} + \mathbf{n}$ after channel estimation has been performed and optimal nearest neighbour decoding \cite{Elayach2014}, the channel capacity
$$C= \log_2 \det \big(\mathbf{Id} + \mathbb{E}\left[\mathbf{n}\mathbf{n}^H\right]^{-1}\mathbf{H}\mathbb{E}\left[\mathbf{x}\mathbf{x}^H\right]\mathbf{H}^H\big)$$
should be studied. Given the noise covariance $\mathbb{E}\left[\mathbf{n}\mathbf{n}^H\right] = \sigma^2\mathbf{Id}$ and the channel singular value decomposition (SVD)
$\mathbf{H} = \mathbf{U}\boldsymbol{\Lambda}\mathbf{V}^H$, the optimal signal covariance is given by $\mathbb{E}\left[\mathbf{x}\mathbf{x}^H\right]_\text{opt} = \mathbf{VDV}^H$, where the diagonal matrix $\mathbf{D}$ corresponds to the optimal power allocation (computed by water-filling). Let us assume here that the channel is fixed but unknown to the system, which uses an estimate $\hat{\mathbf{H}}$ instead (see \cite{Roh2006} for a similar model). Moreover, given the SVD $\hat{\mathbf{H}} = \hat{\mathbf{U}}\hat{\boldsymbol{\Lambda}}\hat{\mathbf{V}}^H$, assume that the signal covariance is taken according to the imperfect estimate as
$\mathbb{E}\left[\mathbf{x}\mathbf{x}^H\right] = \hat{\mathbf{V}}\hat{\mathbf{D}}\hat{\mathbf{V}}^H$, leading to a mismatched precoder.

In that case, $C = C_\text{opt} - C_\text{loss}$, with
$$C_\text{opt} = \log_2 \det \Big(\mathbf{Id} + \tfrac{1}{\sigma^2}\boldsymbol{\Lambda}\mathbf{D}\boldsymbol{\Lambda}^H\Big)$$
and
$$C_\text{loss} = -\log_2 \det \left(\mathbf{Id} + \mathbf{A}\Big(\boldsymbol{\Lambda} \big( \mathbf{V}^H\hat{\mathbf{V}}\hat{\mathbf{D}}\hat{\mathbf{V}}^H\mathbf{V} - \mathbf{D} \big)\boldsymbol{\Lambda}\Big)\right)$$
with $\mathbf{A} = \frac{1}{\sigma^2}\big(\mathbf{Id} + \frac{1}{\sigma^2}\boldsymbol{\Lambda}\mathbf{D}\boldsymbol{\Lambda}^H\big)^{-1}$ (see \cite{Roh2006} for more details). This expression is difficult to study in the general case. However, it simplifies if the receiver has only one antenna ($N_r=1$), to
$$C_\text{opt} = \log_2 \left(1+\text{pSNR}\right)$$
and
%\begin{equation}
%C_\text{loss} = -\log_2\Bigg( 
%1 + \text{pSNR}\frac{\frac{|\langle\mathbf{h},\hat{\mathbf{h}} \rangle|^2}{\Vert\mathbf{h} \Vert_2^2\Vert\hat{\mathbf{h}} \Vert_2^2}-1}{1+\text{pSNR}} \Bigg).
%\label{eq:capacity_loss}
%\end{equation}
\begin{equation}
C_\text{loss} = C_\text{opt}-\log_2 \left(1+\tfrac{|\langle\mathbf{h},\hat{\mathbf{h}} \rangle|^2}{\Vert\mathbf{h} \Vert_2^2\Vert\hat{\mathbf{h}} \Vert_2^2}\text{pSNR}\right).
\label{eq:capacity_loss}
\end{equation}
This capacity loss can be linked to the rMSE of channel estimation by the following bound:
%\begin{equation}
%C_\text{loss} \leq - \log_2 \Bigg( 
%1 + \text{pSNR} \frac{\text{rMSE}(\hat{\mathbf{h}})\big(\text{rMSE}(\hat{\mathbf{h}}) - 2 \big)}{1+\text{pSNR}}
%\Bigg),
%\label{eq:bound_capacity_loss}
%\end{equation}
\begin{equation}
C_\text{loss} \leq C_\text{opt}-\log_2 \left(1+\text{pSNR}\big(1-\text{rMSE}(\hat{\mathbf{h}})\big[2-\text{rMSE}(\hat{\mathbf{h}})\big]\big)\right),
\label{eq:bound_capacity_loss}
\end{equation}
proven in appendix~\ref{app:proof_bound}. Note that the bound is valid as soon as $\text{rMSE}\leq 1$. This bound gives a worst-case data rate loss depending only on the rMSE and the pSNR. It is used in section~\ref{sec:experiments} to assess this loss for realistic channels.

\section{Experiments}
\label{sec:experiments}
The objective of this section is to assess empirically the mathematical developments of the previous sections by: determining if using the oracle estimator makes sense, comparing physical channel estimators to others and quantifying the data rate loss caused by physical channel estimation. All experiments performed here are done using realistic channels generated with help of the NYUSIM channel simulator \cite{Sun2017}, in a millimeter wave massive MIMO downlink context. In particular, the frequency is set to $f=28$\,GHz and the distance between transmitter and receiver to $d=30$\,m to obtain the DoDs, DoAs, gains and phases of each path. The channel matrix is then obtained from \eqref{eq:channel_model_phys} (with the total number of physical paths $P$ between fifty and a hundred), considering a square uniform planar array (UPA) with half-wavelength separated antennas with $N_t = 64$ (unless otherwise stated) at the transmitter and a single antenna receiver ($N_r = 1$). All results shown here are averages over one hundred channel realizations.

\begin{figure}[htbp]
\includegraphics[width=\columnwidth]{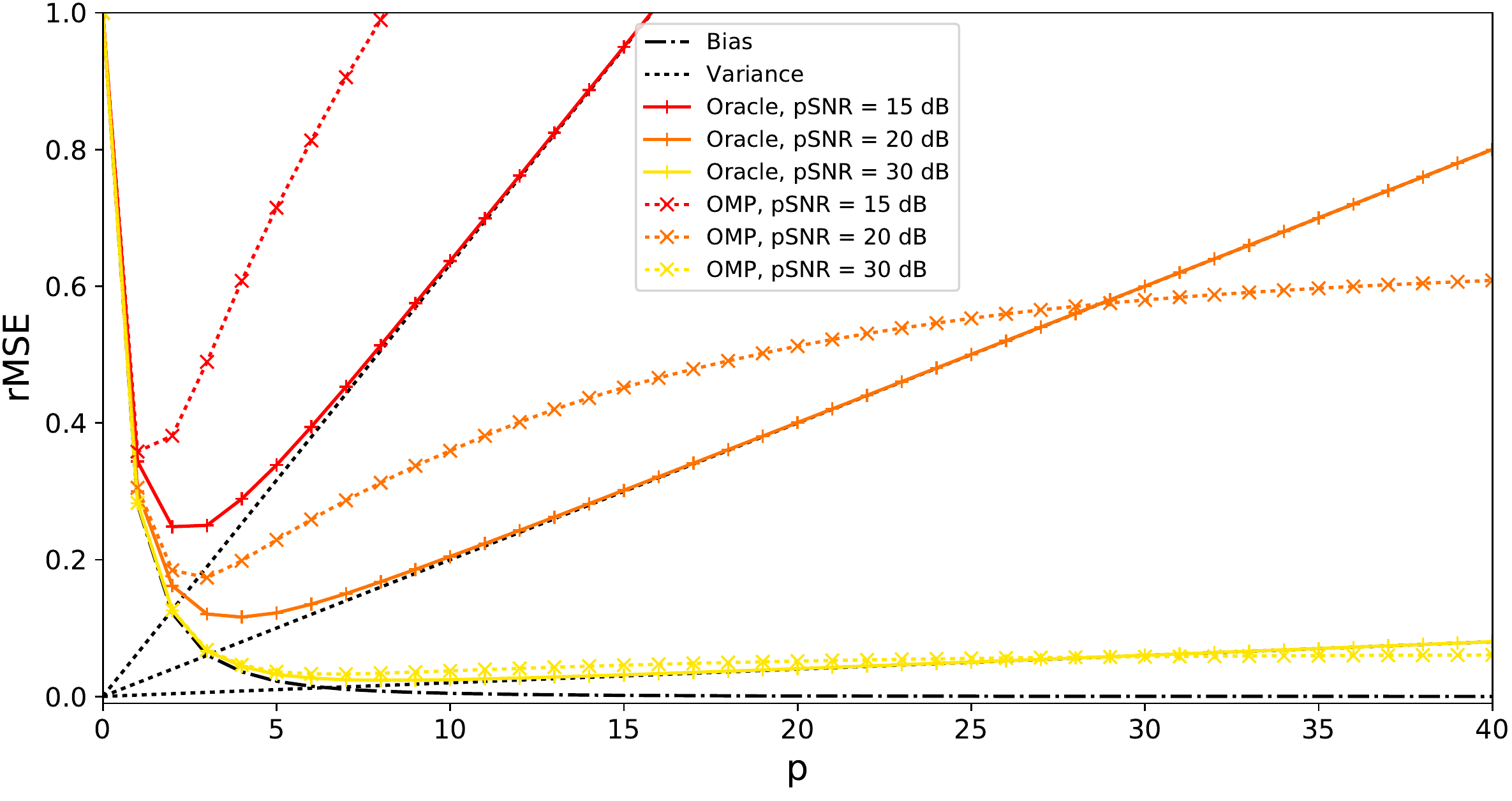}
\caption{Oracle compared to the OMP algorithm for various pSNRs, with respect to $p$.}
\label{fig:oracle_relevance}
\end{figure}

\noindent {\bf Relevance of the oracle.} First, the goal is to determine whether or not the estimator $\hat{\mathbf{h}}_{\text{oracle}}$ has a behavior close to practical estimators. To do so, the bias of the oracle \eqref{eq:bias_oracle} is numerically evaluated by computing $\text{proj}_{\mathcal{M}_p}(\mathbf{h})$ with help of the orthogonal matching pursuit algorithm (OMP) \cite{Tropp2007}, called directly on $\mathbf{h}$ (without noise). The variance of $\hat{\mathbf{h}}_{\text{oracle}}$ is taken directly as \eqref{eq:thm_variance} with an equality (assuming optimal observations), and its rMSE is obtained by summing bias and variance. The oracle is compared to a practical estimator taking as input noisy observations of $\mathbf{h}$, as in \eqref{eq:observation_model_vect} with an optimal observation matrix, using OMP to get channel estimates. Results in terms of rMSE with respect to the number of virtual paths $p$ for various pSNRs are shown on figure~\ref{fig:oracle_relevance}, in which the bias and variances of the oracle are also plotted (in black) to enhance clarity. 
First, one can notice that the oracle and the practical estimator exhibit a similar bias variance tradeoff, especially at high pSNR (note that for $64$ antennas and no precoding, $\text{pSNR} = \text{SNR} + 18$\,dB). This allows to confirm that the oracle estimator is relevant, because it allows a separate study of the bias and variance while providing accurate performance predictions. Note that for a high $p$, the OMP estimator leads to a lower rMSE than the oracle, this is because it yields biased estimates with respect to the projection onto the model (thus the bound on the variance does not apply). Finally, note that the optimal number of virtual paths $p$ yielding the smallest rMSE is very small compared to the number of physical paths $P$, (no more than a dozen against more than fifty), showing empirically that physical paths can indeed be merged into fewer virtual paths with little accuracy loss.

\begin{figure*}[htbp]
\begin{subfigure}[b]{0.33\textwidth}
\includegraphics[width=\columnwidth]{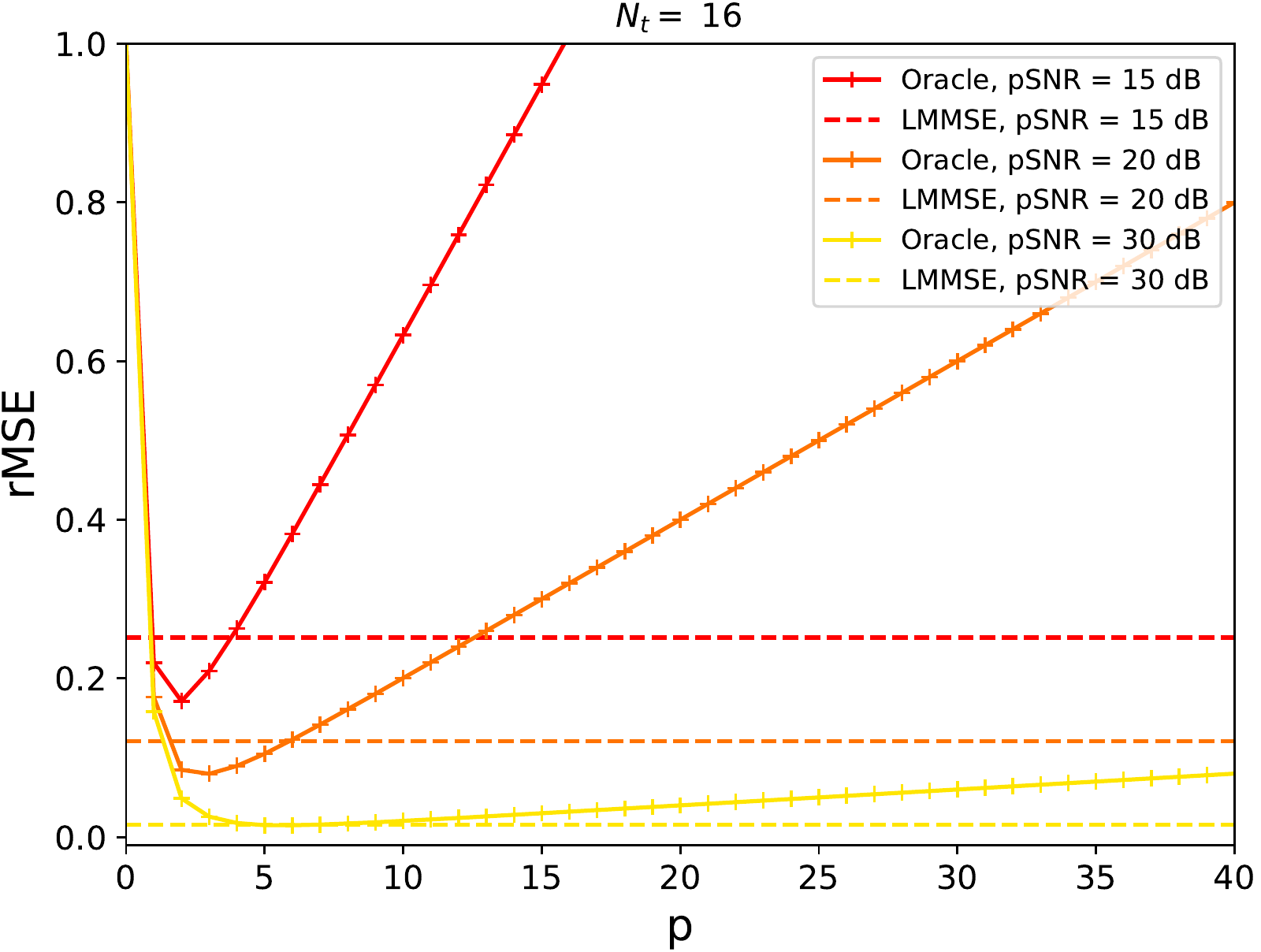}
\end{subfigure}
\begin{subfigure}[b]{0.33\textwidth}
\includegraphics[width=\columnwidth]{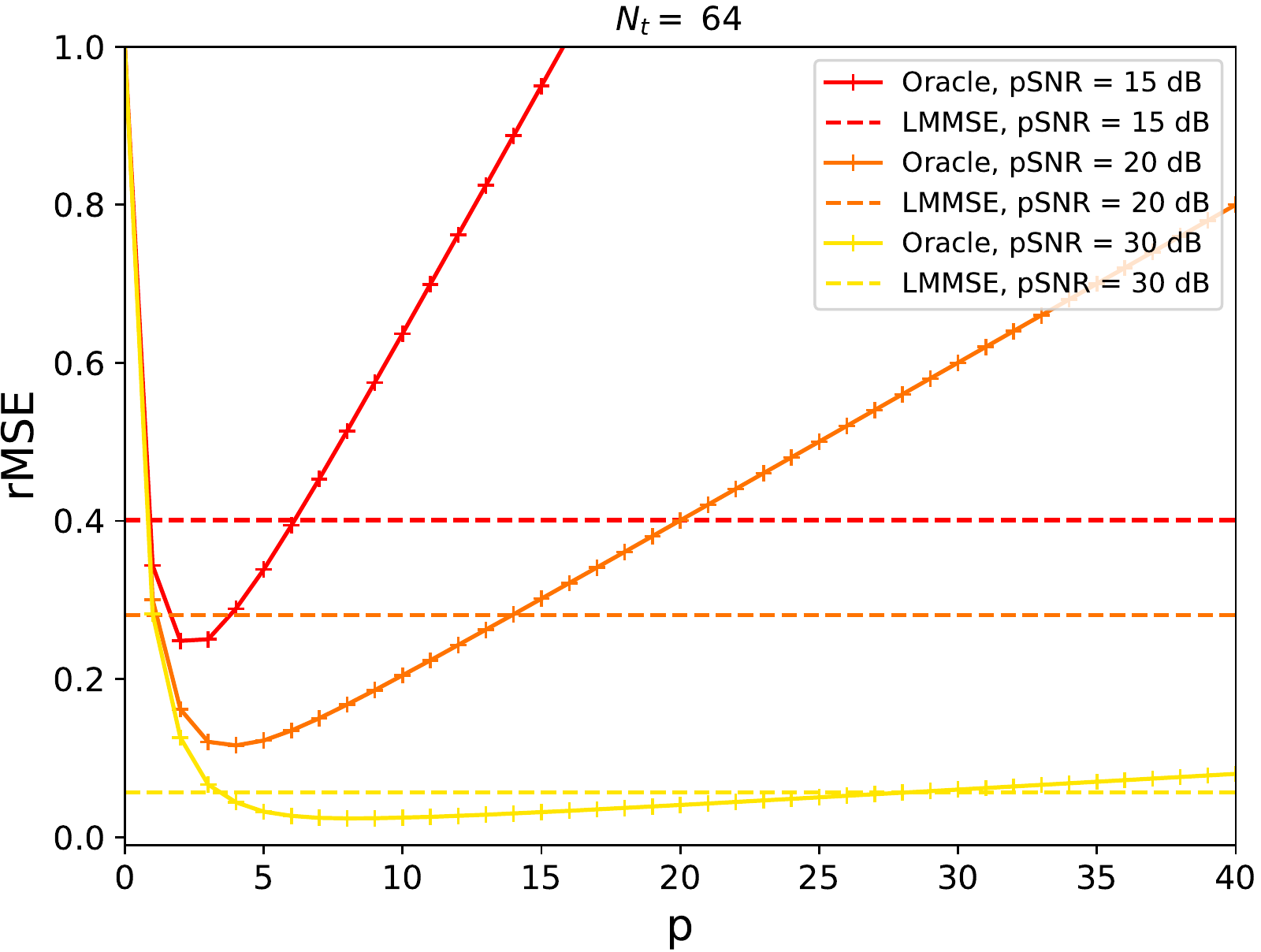}
\end{subfigure}
\begin{subfigure}[b]{0.33\textwidth}
\includegraphics[width=\columnwidth]{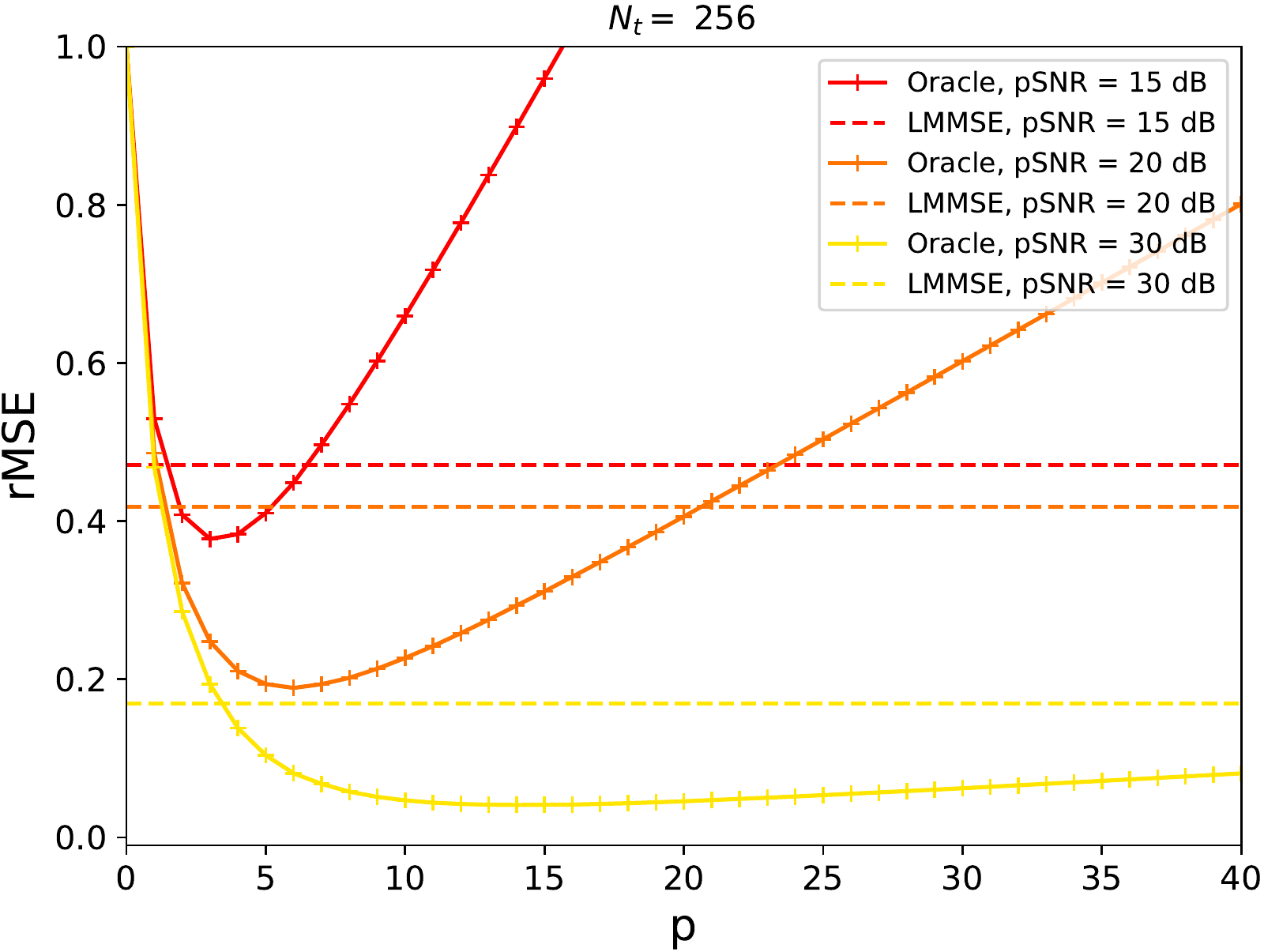}
\end{subfigure}
\caption{Comparison of the oracle with the LMMSE for various numbers of transmit antennas and pSNRs, with respect to $p$.}
\label{fig:comp_LMMSE}
\end{figure*}

\noindent {\bf Comparison with LMMSE.} Let us now compare the oracle to other estimators presented in section~\ref{ssec:classical_estimators}. Actually, the comparison is performed here with the LMMSE only, the comparison with the LS estimator not being shown for readability and brevity reasons. This is justified by the fact that the LS is always worse than the LMMSE in terms of optimal rMSE, but the same trends are observed, due to the similar expressions of the optimal performance \eqref{eq:perf_LS} and \eqref{eq:perf_LMMSE}, leading to the same conclusions. The performance of LMMSE is taken according to \eqref{eq:perf_LMMSE}, assuming the covariance matrix $\mathbf{R}$ is perfectly known, and taking $\frac{\text{Tr}\left[\mathbf{R}^{-1}\right]\text{Tr}\left[\mathbf{R}\right]}{(N_rN_t)^2}=2$ (this choice is rather arbitrary, but the point here is to show the relative behavior of the methods, assessing this quantity in practice would of course be valuable but is out of the scope of the present paper). Results are shown on figure~\ref{fig:comp_LMMSE} for various number of transmit antennas. One important thing to notice is that especially at high pSNR, having a lot of transmit antennas degrades the performance of LMMSE more than it degrades the performance of the oracle estimator. This indicates that for massive MIMO with a lot of antennas, physical channel estimation methods may be better suited than LMMSE. Second, one can see that the optimal number of virtual paths to consider for the oracle estimator grows with the number of antennas. This behavior was expected from the bound on the bias given in section~\ref{sec:bound_bias}, since the quantity $\kappa(\mathcal{A})$ grows with $N_t$ (the angular resolution increases). Finally, note that even though the curves indicate that the estimation task is harder for a greater number of antennas at constant pSNR, high pSNRs are easier to attain with a lot of antennas since according to \eqref{eq:pSNR}, pSNR is proportional to $N_t$.

\begin{figure}[htbp]
\includegraphics[width=\columnwidth]{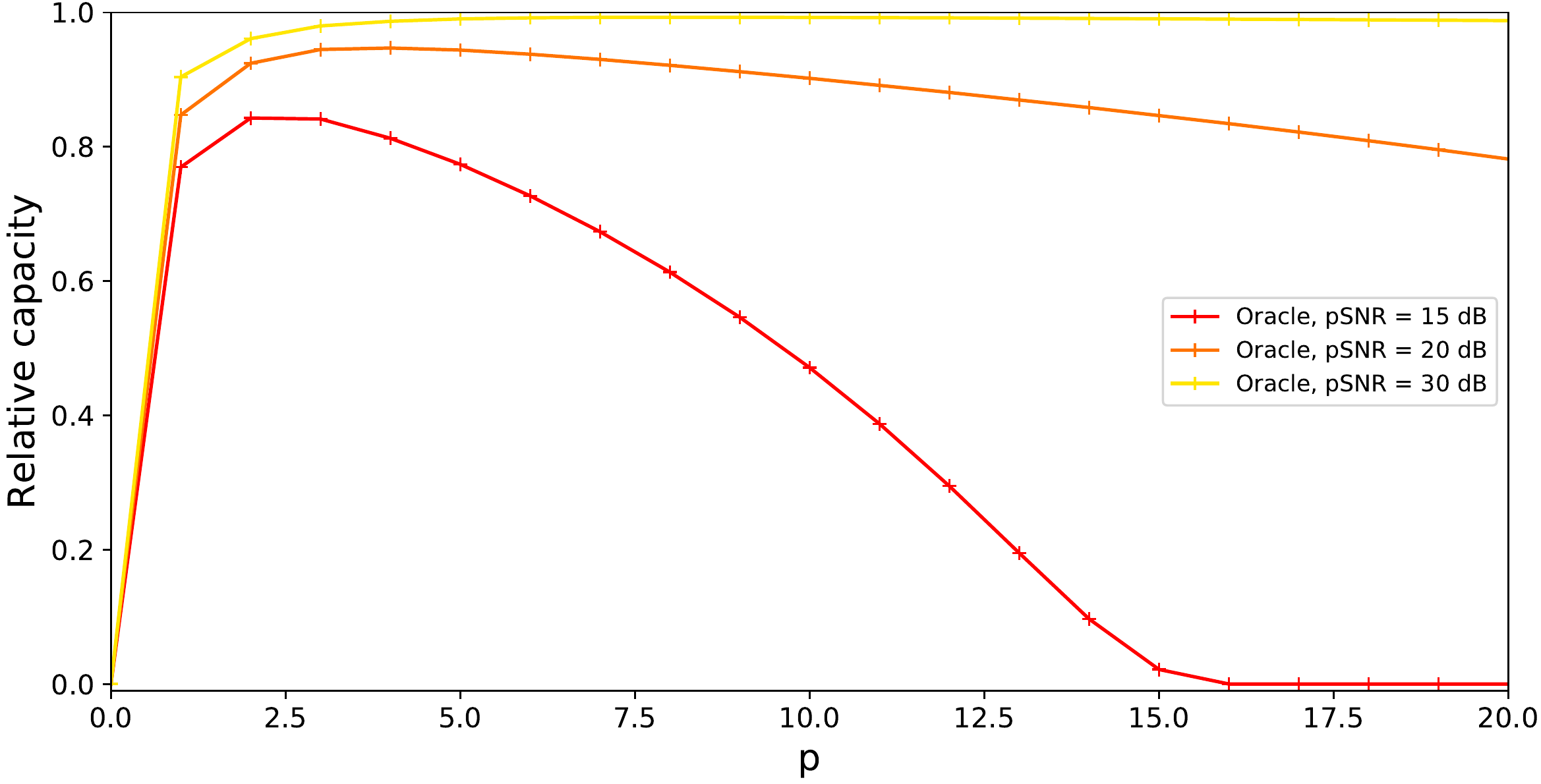}
\caption{Relative capacity with respect to $p$ for various pSNRs.}
\label{fig:capacity_loss}
\end{figure} 
\noindent {\bf Impact on data rate.} Finally, let us quantify the data rate loss caused by the estimation error when using a physical model, with respect to the number of considered virtual paths. To do so, the bound \eqref{eq:capacity_loss} is used with the rMSE of the oracle estimator, the results being shown on figure~\ref{fig:capacity_loss}, where the relative capacity is defined as $\frac{C_\text{opt} - C_\text{loss}}{C_\text{opt}}$. These curves show it is sufficient to estimate very few virtual paths to guarantee a data rate close to the optimal one (obtained if the channel is perfectly known) in the tested millimeter wave context. For example, with a pSNR of $20$\,dB, only four virtual paths are sufficient to attain $94$\% of the optimal capacity. The higher the pSNR, the higher the optimal (in terms of capacity) number of virtual paths to estimate, which makes sense, since more physical paths are then above the noise level.

\section{Conclusion}
In this paper, the performance of MIMO channel estimators using a physical model was theoretically studied. To do so, an appropriate oracle estimator was defined which allowed to study separately the bias and variance terms of the rMSE. The variance term having been already studied in a previous work \cite{Lemagoarou2018}, the present paper focused on the bias term, which was bounded by an interpretable quantity depending both on the propagation properties of the channel and on the antenna arrays geometries. Moreover, the data rate loss caused by the channel estimation error was bounded by a quantity depending only on the rMSE and the pSNR. Then, the defined oracle, whose relevance was empirically confirmed on realistic channels in a millimeter wave massive MIMO context, was compared to the classical LMMSE estimator, showing that channel estimators based on a physical model may be better suited to systems with a large number of antennas. Finally, it was shown empirically that estimating very few virtual paths is sufficient to guarantee a data rate close to the optimum in the aforementioned context.

In the future, it would be of great interest to perform extensive comparisons between physical channel estimators and other types of estimators such as the LMMSE, with the objective to see emerge different characteristic settings in which one or the other type of estimator is more adapted. 
Such a study would be very useful for system design. On a more technical side, improving the bound on the bias proposed here should be possible, as mentioned in appendix~\ref{app:improving_the_bound}.

%\appendices
%\titleformat{\section}[display]
%  {\normalfont\Large\bfseries}{\appendixname\enspace\thesection}{.5em}{}
\appendix
\subsection{Proof of lemma~\ref{lem:approx_scalar_product}}
\label{app:proof_lemma}
Using the series representation of the exponential $
\mathrm{e}^x = \sum_{l=0}^{+\infty}\frac{x^l}{l!}
$, the scalar product is expressed
$$
\mathbf{e}(\overrightarrow{v_{k}})^H\mathbf{e}(\overrightarrow{u_{i}}) = \frac{1}{N}\sum_{n=1}^N\sum_{l=0}^{+\infty}\frac{\big(-\mathrm{j}\frac{2\pi}{\lambda}\overrightarrow{a_n}.(\overrightarrow{u_i} - \overrightarrow{v_k})\big)^l }{l!}.
$$
Using $\sum_{n=1}^{N}\overrightarrow{a_n}=0$ since the antennas are located with respect to the centroid of the array, the term $l=1$ is null, leading to
$$
\mathbf{e}(\overrightarrow{v_{k}})^H\mathbf{e}(\overrightarrow{u_{i}}) = 1+\sum_{l=2}^{+\infty}\frac{\big(-\mathrm{j}2\pi\big)^l }{l!}\frac{1}{N}\sum_{n=1}^N\big(\frac{1}{\lambda}\overrightarrow{a_n}.(\overrightarrow{u_i} - \overrightarrow{v_k})\big)^l. 
$$
Both the real and imaginary parts of this series are alternating series. The terms are decreasing in magnitude if
$$
\frac{1}{l!}\left| \frac{2\pi}{\lambda}\overrightarrow{a_n}.(\overrightarrow{u_i} - \overrightarrow{v_k}) \right|^l > \frac{1}{(l+2)!}\left| \frac{2\pi}{\lambda}\overrightarrow{a_n}.(\overrightarrow{u_i} - \overrightarrow{v_k}) \right|^{(l+2)},\,\forall n. 
$$
This condition is fulfilled as soon as $$\left\Vert \overrightarrow{u_i} - \overrightarrow{v_k} \right\Vert_2 < \frac{1}{\sqrt{2}\pi\left\Vert \frac{\overrightarrow{a_n}}{\lambda} \right\Vert_2 },\, \forall n, $$ (taking $l=0$ and applying the Cauchy-Schwarz inequality). In that case, $\mathfrak{Re}\left\{ \mathbf{e}(\overrightarrow{v_{k}})^H\mathbf{e}(\overrightarrow{u_{i}}) \right\} \geq 1-2\pi^2\frac{1}{N}\sum_{n=1}^N\big(\frac{1}{\lambda}\overrightarrow{a_n}.(\overrightarrow{u_i} - \overrightarrow{v_k})\big)^2$. Using $1- \left| \mathbf{e}(\overrightarrow{v_{k}})^H\mathbf{e}(\overrightarrow{u_{i}}) \right|^2 \leq 1-\mathfrak{Re}\left\{ \mathbf{e}(\overrightarrow{v_{k}})^H\mathbf{e}(\overrightarrow{u_{i}}) \right\}^2$ and $\mathfrak{Re}\left\{ \mathbf{e}(\overrightarrow{v_{k}})^H\mathbf{e}(\overrightarrow{u_{i}}) \right\}^2 \geq 1-4\pi^2\frac{1}{N}\sum_{n=1}^N\big(\frac{1}{\lambda}\overrightarrow{a_n}.(\overrightarrow{u_i} - \overrightarrow{v_k})\big)^2$,
$$
1- \left| \mathbf{e}(\overrightarrow{v_{k}})^H\mathbf{e}(\overrightarrow{u_{i}}) \right|^2 \leq 4\pi^2\frac{1}{N}\sum\nolimits_{n=1}^N\big(\frac{1}{\lambda}\overrightarrow{a_n}.(\overrightarrow{u_i} - \overrightarrow{v_k})\big)^2.
$$
Developing the scalar product %and taking the square root 
proves the lemma.
%$$
%\mathbf{e}(\overrightarrow{v_{k}})^H\mathbf{e}(\overrightarrow{u_{i}}) = 1+\sum_{l=2}^{+\infty}\frac{\big(-\mathrm{j}2\pi\big)^l }{l!}\left\Vert\overrightarrow{u_i} - \overrightarrow{v_k} \right\Vert_2^l\frac{1}{N}\sum_{n=1}^N\left\Vert \frac{\overrightarrow{a_n}}{\lambda} \right\Vert_2^l\cos\big(\overrightarrow{a_n},(\overrightarrow{u_i} - \overrightarrow{v_k})\big)^l 
%$$

\subsection{Improving the bound}
\label{app:improving_the_bound}
The bound of this paper is obtained applying the triangle inequality twice in a row, which may lead to a loose bound. Instead, one could apply it only once to get
$$
\big\Vert\mathbf{h}-\hat{\mathbf{h}}\big\Vert_2\leq 
\sum\limits_{k=1}^p\Big\Vert\sum\limits_{i\in \mathcal{R}_k} c_i\left(\mathbf{e}(\overrightarrow{u_{i}}) - \mathbf{e}(\overrightarrow{v_{k}})^H\mathbf{e}(\overrightarrow{u_{i}})\mathbf{e}(\overrightarrow{v_{k}})\right)\Big\Vert_2.
$$
At this point, one can notice that each term of the sum, when squared, corresponds to a quadratic form 
$$
\Big\Vert\sum\limits_{i\in \mathcal{R}_k} c_i\left(\mathbf{e}(\overrightarrow{u_{i}}) - \mathbf{e}(\overrightarrow{v_{k}})^H\mathbf{e}(\overrightarrow{u_{i}})\mathbf{e}(\overrightarrow{v_{k}})\right)\Big\Vert_2^2 = \mathbf{c}^H\mathbf{Q}\mathbf{c},
$$
where $\mathbf{c} = (c_1,\dots,c_{|\mathcal{R}_k|})^T$ and $\mathbf{Q} \in \mathbb{C}^{|\mathcal{R}_k|\times |\mathcal{R}_k|}$
with
$$
q_{ij} = \mathbf{e}(\overrightarrow{u_{i}})^H\mathbf{e}(\overrightarrow{u_{j}}) - \mathbf{e}(\overrightarrow{u_{i}})^H\mathbf{e}(\overrightarrow{v_{k}})\mathbf{e}(\overrightarrow{v_{k}})^H\mathbf{e}(\overrightarrow{u_{j}}).
$$
Studying the properties of the matrix $\mathbf{Q}$ may lead to a tighter bound, but is not guaranteed to lead to interpretable results.
\subsection{Proof of \eqref{eq:bound_capacity_loss}}
\label{app:proof_bound}
Starting from \eqref{eq:capacity_loss}, one can first notice that
$\frac{|\langle\mathbf{h},\hat{\mathbf{h}} \rangle|^2}{\Vert\mathbf{h} \Vert_2^2\Vert\hat{\mathbf{h}} \Vert_2^2} \geq \frac{\mathfrak{Re}(\langle\mathbf{h},\hat{\mathbf{h}} \rangle)^2}{\Vert\mathbf{h} \Vert_2^2\Vert\hat{\mathbf{h}} \Vert_2^2}$, with
$$
\frac{\mathfrak{Re}(\langle\mathbf{h},\hat{\mathbf{h}} \rangle)^2}{\Vert\mathbf{h} \Vert_2^2\Vert\hat{\mathbf{h}} \Vert_2^2} = 1 + \tfrac{1}{4}\Big\Vert \frac{\mathbf{h}}{\left\Vert \mathbf{h} \right\Vert_2} - \frac{\hat{\mathbf{h}}}{\Vert \hat{\mathbf{h}} \Vert_2} \Big\Vert_2^4 - \Big\Vert \frac{\mathbf{h}}{\left\Vert \mathbf{h} \right\Vert_2} - \frac{\hat{\mathbf{h}}}{\Vert \hat{\mathbf{h}} \Vert_2} \Big\Vert_2^2.$$ Then, writing $ z\triangleq
\Big\Vert \frac{\mathbf{h}}{\left\Vert \mathbf{h} \right\Vert_2} - \frac{\hat{\mathbf{h}}}{\Vert \hat{\mathbf{h}} \Vert_2} \Big\Vert_2^2
=
2\frac{\Vert \mathbf{h} - \hat{\mathbf{h}} \Vert_2^2}{\left\Vert \mathbf{h} \right\Vert_2^2} -2x^2 + 4\alpha x - 2\alpha
$ with $x \triangleq \frac{\Vert \hat{\mathbf{h}} \Vert_2}{\Vert \mathbf{h} \Vert_2}$ and $\alpha \triangleq \frac{\mathfrak{Re}(\langle\mathbf{h},\hat{\mathbf{h}} \rangle)}{\Vert\mathbf{h} \Vert_2\Vert\hat{\mathbf{h}} \Vert_2}$, the bound
$$
z=\Big\Vert \frac{\mathbf{h}}{\left\Vert \mathbf{h} \right\Vert_2} - \frac{\hat{\mathbf{h}}}{\Vert \hat{\mathbf{h}} \Vert_2} \Big\Vert_2^2
\leq
2\frac{\Vert \mathbf{h} - \hat{\mathbf{h}} \Vert_2^2}{\left\Vert \mathbf{h} \right\Vert_2^2} = 2\text{rMSE}
$$
holds if $\text{rMSE}\leq 1$. Indeed, in that case $\mathfrak{Re}(\langle\mathbf{h},\hat{\mathbf{h}} \rangle)\geq 0$ which yields  $-2x^2 + 4\alpha x - 2\alpha \leq 0$. The function $1+\frac{z^2}{4}-z$ decreases for $z\leq 2$, which finishes the proof yielding
$$
\frac{|\langle\mathbf{h},\hat{\mathbf{h}} \rangle|^2}{\Vert\mathbf{h} \Vert_2^2\Vert\hat{\mathbf{h}} \Vert_2^2} \geq 1-\text{rMSE}(2-\text{rMSE}).
$$

\bibliographystyle{IEEEbib}
\bibliography{biblio_mimo}

\end{document}